\documentclass[12pt]{article}

\usepackage{amsmath}
\usepackage{amssymb}
\usepackage{amsfonts}
\usepackage{amscd}
\usepackage{amsbsy}
\usepackage{amsthm}
\usepackage{latexsym}
\usepackage{graphicx,color,epsf}
\usepackage{slashed}

\def\nn{\nonumber}       
\def\n{\label}                 
\def\r{\ref}                    

\newcommand{\eq}[1]{(\ref{#1})}

\def\beq{\begin{eqnarray}}
\def\eeq{\end{eqnarray}}
\def\ln{\,\mbox{ln}\,}

\def\Det{\,\mbox{Det}\,}

\def\diag{\,\mbox{diag}\,}
\def\Tr{\,\mbox{Tr}\,}
\def\sTr{\,\mbox{sTr}\,}

\def\al{\alpha}
\def\be{\beta}
\def\ch{\chi}
\def\ga{\gamma}
\def\de{\delta}
\def\vp{\varepsilon}
\def\ep{\epsilon}
\def\ze{\zeta}

\def\la{\lambda}

\def\pa{\partial}

\def\si{\sigma}

\def\ph{\varphi}

\def\Ga{\Gamma}

\def\Om{\Omega}

\def\Th{\Theta}

\begin{document}

\begin{center}

{\Large \bf Quantum aspects of Yukawa model with scalar and axial
scalar fields in curved spacetime} \vskip 4mm

{\bf Iosif L. Buchbinder}$^{b,c}$\footnote{E-mail
address: \ joseph@tspu.edu.ru},
\  \
{\bf Andreza Rairis Rodrigues}$^{a}$\footnote{E-mail address: \ andrezarodrigues@ice.ufjf.br},
\\
{\bf Eduardo Antonio dos Reis}$^{a}$\footnote{E-mail address: \ eduardoreis@ice.ufjf.br}
\ \ and
\  \ {\bf Ilya L. Shapiro}$^{a,b,c}$\footnote{E-mail address: \
shapiro@fisica.ufjf.br}.
\vskip 4mm

$^a$ \
{\sl
Departamento de F\'{\i}sica, \ ICE, \ Universidade Federal de Juiz de Fora
\\
36036-330, Juiz de Fora, \ MG, \ Brazil}
\vskip 2mm

$^b$ \
{\sl
Department of Theoretical Physics, Tomsk State Pedagogical
University
\\
634061, Tomsk, Russia}
\vskip 2mm

$^c$ \
{\sl
National Research Tomsk State University, 634050, Tomsk, Russia}
\end{center}

\vskip 4mm


\begin{quotation}
\noindent
{\bf Abstract.} \\
We study the Yukawa model with one scalar and one axial scalar
fields, coupled to $N$ copies of Dirac fermions, in curved spacetime
background. The theory possesses a reach set of coupling constants,
including the scalar terms with odd powers of scalar fields in the
potential, and constants of non-minimal coupling of the scalar
fields to gravity. Using the heat-kernel technique and dimensional
regularization, we derive the one-loop divergences, describe the
renormalization of the theory under consideration and calculate the
full set of beta- and gamma-functions for all coupling constants and
fields. As a next step, we construct the renormalized one-loop
effective potential of the scalar fields up to the terms linear in
scalar curvature. This calculation includes only the contributions
from quantum scalar fields, and is performed using covariant cut-off
regularization and local momentum representation. Some difficulties
of the renormalization group approach to the effective potential in
the case under consideration are discussed.


\vskip 4mm

\end{quotation}

\section{Introduction}
\label{int}

The interaction between scalar fields with Dirac spinors through a
Yukawa interaction is attracting a special attention in quantum field
theory in curved spacetime. In this respect one can mention recent
analysis that includes both scalar and a pseudoscalar couplings
\cite{TomsJHEP} and more recently with the inclusion of a gauge
field \cite{TomsPRD,Toms2019}. In the present work we continue
the previous treatment of Yukawa model with sterile scalar discussed
in \cite{Barra} and extend it to the case of the two (ordinary and
axial) scalars with a Yukawa coupling to fermions and general
renormalizable form of self-interaction.
Our immediate purpose will be the calculation of divergences in the
most economic way, as it was done in the original publication on
the renormalization of the Abelian model with Yukawa coupling in
curved spacetime with torsion from long ago \cite{BuchSha-90}
(see also the book \cite{book}) .

Similar consideration of the simpler model with a single scalar field
was useful in establishing the constraints on the quantum theory that
come from the condition of renormalizability of the Abelian theory
with massive Dirac field. The form of the self-interaction potential
of a scalar field ensuring the renormalizability of such a theory is
an interesting aspect, that was not explored completely in the original
work \cite{BuchSha-90}. It was shown and discussed in details in
the recent work \cite{Barra} that the renormalizable scalar with
Yukawa interaction includes self-interactions with {\it odd} powers
of the scalar fields. These qualitatively new interactions include
linear term, the term with a cubic coupling, and also a linear term
describing the interaction between scalar field and scalar curvature.

In all examples of renormalizable quantum field theories with scalar
fields, which were known until now, it was possible to construct
solutions to the renormalization group equation for the effective
action which enable to derive the effective potential in the most
economic way \cite{ColeWein},  including in curved spacetime
\cite{BuchOd84} (see also the  generalization to other sectors of
effective action in \cite{BW,book}). In the model with a single
sterile scalar one has to extend this nicely working scheme to
include odd powers of the scalar field, with this generalization it
still works pretty well \cite{Barra}.

The generalization of these considerations to the parity-preserving
model with an additional axial scalar field is an interesting and
challenging problem. Let us start by stating that this problem
makes sense from the viewpoint of physical applications. First
of all, there is an important example of an axial scalar, that is an
axion. Regardless axion might have different form
of coupling to gauge and fermion fields compared to an ordinary
axial scalar field, it is interesting to explore the renormalization
of such a parity-preserving model on a simple example. On the
other hand, in the recent years there were indications of the
possible violations of parity in the gravitational action as an
explanation of some astrophysical observations \cite{ASNY}.
Therefore it may be interesting to have a consistent description of
the models which are capable to explain such a violation, and the
study of renormalization of the model with axial scalar may be a
useful step in better understanding of a possible quantum origin
of such terms.

Another interesting aspect of the model under consideration is
that such a theory has two different scalar masses, that is a usual
situation in effective field theory (see e.g. the book \cite{Ilisie}).
In the recent work \cite{324} we explored the non-local finite
contributions of the curved-spacetime diagrams with mixed
internal lines, e.g. one of a light and another of a heavy scalar
fields. Here we supplement this result by deriving the effective
potential in the two-scalar model.
It is worth pointing out that this situation is typical for effective
field theories, especially the ones with different mass scales and
diagrams with mixed types of internal lines. The effective potential
involves two independent contributions, one from the loops of
scalar fields and another one from the spinor loop. In what follows
we show that the results for these contributions look somehow
unusual. In the scalar sector we meet a complicated non-polynomial
mixing of the scalar masses and couplings,
something one could expect for the two-scalar model.

The paper is organized as follows. In Sec.~\ref{s2} we describe
the model including a real scalar field and a pseudoscalar field
coupled to $N$-component fermionic field and derive the
corresponding one-loop divergences. The one-loop renormalization
relations in this theory and the derivation of the renormalization
group functions are collected in Sec.~\ref{s3}. In Sec.~\ref{s4}
the renormalized one-loop effective potential is derived by using
the local momentum representation. Finally, our conclusions and
the discussion of the results are presented in Sec.~\ref{s6}.

\section{Yukawa model and its renormalization}
\label{s2}

Consider a Yukawa model including a real scalar field $\,\ph\,$ and a
real pseudoscalar (axial scalar) field $\,\ch$, coupled to the $N$
copies of a fermionic field $\,{\Psi}_i$, with the classical action of
the form
\beq
S &=&
\int d^{4} x \sqrt{-g} \Big\{
 \bar{\Psi}_i  \left(  i \slashed{\nabla}
- M - h_1\varphi - h_2\ch \gamma^{5} \right)\delta^{ij}  \Psi_j
+ \frac12 g^{\mu\nu}\partial_{\mu}\varphi\partial_{\nu}\varphi
\nn
\\
&+&
\frac12 g^{\mu\nu}\partial_{\mu}\ch\partial_{\nu}\ch
- \frac12 m^{2}_1 \varphi^{2}  - \frac12 m^{2}_2 \ch^{2}
+\frac12\xi_1 R \varphi^{2} +\frac12\xi_2 R \ch^{2}
- \frac{ \lambda_1}{4!} \varphi^{4}
- \frac{ \lambda_2}{4!} \ch^{4}
\nn
\\
&-&
\frac{ \lambda_3}{2} \varphi^2 \ch^{2} - \frac{g}{3!} \varphi^{3}
- \frac{p}{2} \varphi \ch^2
- \tau \varphi - fR \ph
\Big\},
\n{1act}
\label{classact}
\eeq
where $m_1, m_2$ and $M$ are  respectively the masses of scalar,
pseudoscalar and spinor fields, $h_1$ and $h_2$ are the Yukawa
coupling constants. Finally, $\la_1$, $\la_2$, $\la_3,$ $g,$ $p$ and
$\tau$  are coupling constants in the scalar -- pseudoscalar sectors,
that survive in the flat limit, while $\xi_1$  and $f$  are the
nonminimal parameters of the scalar field and $\xi_2$ the nonminimal
parameter of the interaction between axial scalar field with gravity.
It is easy to note that the action has not only the standard even terms,
but also a set of odd terms, with the dimensional parameters $g,$
$p$ and $f$. As we shall see in brief, these terms are necessary to
achieve renormalizability of the theory. The last observation is that
term which are linear and cubic in the pseudoscalar field are
excluded by the requirement that the Lagrangian is a parity-even
scalar.

In order to calculate the one-loop divergences, we shall use the
heat-kernel method, and perform the background-quantum
splitting of the fields, according to
\beq
\ph \,\,\rightarrow\,\, \ph + \sigma ,
\quad
\ch \,\,\rightarrow\,\, \ch + \rho,
\quad
\bar{\Psi}_i  \,\,\rightarrow\,\, \bar{\Psi}_i + \bar{\eta}_i,
\quad
\Psi_j  \,\,\rightarrow\,\, \Psi_j + \eta_j \,,
\eeq
where $\ph, \, \ch, \,\bar{\Psi}, \,\Psi$
are the classical background fields and
$\si,\, \rho, \, \bar{\eta},\, \eta$ their quantum counterparts.

The bilinear in quantum fields part of the action is written
as follows
\beq
S^{(2)}
&=&
\frac{1}{2} \int d^{4}x \sqrt{-g} \,\,\, \Big( \sigma
\quad \rho \quad \bar{\eta}_i \Big) \,\hat{H} \, \left(
\begin{array}{c}
\si  \\
\rho \\
\eta_j
\end{array}
\right)
\nn
\\
&=&
\frac{1}{2} \int d^{4} x \sqrt{-g} \Big\{ \sigma H_{11} \sigma
+ \rho H_{21} \sigma + \bar{\eta}_i H_{31} \sigma
+ \sigma H_{12} \rho
\nn
\\
&+&
\rho H_{22} \rho
+ \bar{\eta_i} H_{32} \rho + \sigma H_{13} \eta_j
+ \rho H_{23} \eta_j + \bar{\eta}_i H_{33} \eta_j \Big\},
\label{action}
\eeq
where the elements of the matrix operator $\hat{H}$ have the form
\beq
\n{hatH}
H_{11} &=& - \Box - m^{2}_1 + \xi_1 R -g \ph - \lambda_3 \ch^2
- \frac{\lambda_1}{2} \ph^{2} ,
\nn
\\
H_{12} &=& -4\ph \ch \lambda_3 - 2p \ch ,
\qquad
H_{13} \,=\, -2h_1 \bar{\Psi}_j \, ,
\qquad
H_{21} \,=\, - 4\ph \ch \lambda_3 -2p \ch ,
\nn
\\
H_{22} &=& - \Box - m^{2}_2 + \xi_2 R -p \ph
- \lambda_3 \ph^2 - \frac{\lambda_2}{2} \ch^{2} ,
\nn
\\
H_{23} &=& -2h_2 \bar{\Psi}_j \gamma^5 ,
\qquad
H_{31} \,=\, -2h_1 \Psi_i \, ,
\qquad
H_{32} \,=\, -2h_2 \gamma^5 \Psi_i \, ,
\nn
\\
H_{33} &=&  2(i\slashed{\nabla}
-M - h_1 \ph - h_2 \ch \gamma^5) \delta^{ij} .
\eeq
It proves useful to introduce conjugated matrix operator
\beq
\hat{H}^{*} = \left(
\begin{array}{ccc}
-1 & 0 & 0 \\
0 & - 1 & 0 \\
0 & 0 & -\frac{1}{2} (i\slashed{\nabla} + M)
\end{array}
\right).
\eeq

The one-loop quantum contribution to effective action is defined
by the expression $\,\Tr\ln(\hat{H})$. To calculate the divergences
of effective action we will write it as
\beq
\Tr\ln(\hat{H}) \,=\,\Tr\ln(\hat{H}\hat{H}^*)-\Tr\ln(\hat{H}^*).
\eeq
The last term $\Tr\ln \hat{H}^*$ contributes only to the
vacuum  divergences that are known for an arbitrary model
\cite{birdav,book}. Therefore it is sufficient to calculate the
divergences of the product $\hat{H}\hat{H}^{*}$, that has a
standard form,
\beq
{\cal \hat{H}} &=&
\hat{H}\hat{H}^{*}
= \hat{1} \Box  + 2 \hat{h}^{\mu} \nabla_{\mu} + \hat{\Pi}.
\label{H}
\eeq
Hence,
\beq
{\cal{H}}_{11} &=& \Box + m^{2}_1
+ \frac{\lambda_1}{2} \ph^{2} - \xi_1 R + \lambda_3 \ch^2 + g \ph ,
\nn
\\
{\cal{H}}_{12} &=& 2p\ch + 4 \lambda_3 \ph \ch ,
\qquad
{\cal{H}}_{13} \,=\, h_1 \bar{\Psi}_j (i\slashed{\nabla} + M) ,
\nn
\\
{\cal{H}}_{21} &=& 2p\ch + 4 \lambda_3 \ph \ch ,
\qquad
{\cal{H}}_{22}
\,=\, \Box + m^{2}_2 + \frac{\lambda_2}{2} \ch^{2}
- \xi_2 R + \lambda_3 \ph^2 + p \ph ,
\nn
\\
{\cal{H}}_{23} &=&  h_2 \bar{\Psi}_j \ga^5(i\slashed{\nabla} + M) ,
\qquad
{\cal{H}}_{31} \,=\, 2h_1 \Psi_i \, ,
\qquad
{\cal{H}}_{32} \,=\, 2h_2 \gamma^5 \Psi_i \, ,
\nn
\\
{\cal{H}}_{33} &=& \delta^{ij} \big [\Box
- \frac14R + M^2 + h_1 \ph ( i\slashed{\nabla}+ M)
+  h_2\ch \gamma^5 (i\slashed{\nabla}+M)\big] .
\eeq
where we can identify
\beq
h^{\mu}_{13} &=& \frac{i h_1}{2}  \bar{\Psi}_j \gamma^{\mu} ,
\quad
h^{\mu}_{23} \,=\, \frac{i h_2}{2} \bar{\Psi}_j
\ga^5 \ga^\mu ,
\quad
h^{\mu}_{33} \,=\, \frac{i}{2} (h_1 \ph
+ h_2 \ch \gamma^5 ) \gamma^{\mu}\delta^{ij}
\eeq
and
\beq
\Pi_{11}
&=&
m^{2}_1 + \frac{\lambda_1}{2} \ph^{2}
- \xi_1 R + g\ph + \lambda_3 \ch^2 ,
\quad
\Pi_{12} \,=\, 2p \ch + 4 \lambda_3 \ph \ch ,
\nn
\\
\Pi_{13} &=& h_1M \bar{\Psi}_j \, ,
\quad
\Pi_{21} \,=\, 2p \ch + 4 \lambda_3 \ph \ch ,
\nn
\\
\Pi_{22} &=& m^{2}_2 + \frac{\lambda_2}{2} \ch^{2}
- \xi_2 R + p\ph + \lambda_3 \ph^2 ,
\nn
\\
\Pi_{23} &=& h_2M \bar{\Psi}_j \gamma^5 ,
\quad
\Pi_{31} \,=\, 2h_1 \Psi_i \, ,
\quad
\Pi_{32} \,=\, 2h_2 \gamma^5 \Psi_i \, ,
\nn
\\
\Pi_{33} &=& \delta^{ij}\Big[ M^{2} -\frac{1}{4} R
+ h_1M \ph + h_2 M \ch \gamma^5 \Big] .
\eeq

The Schwinger--De-Witt proper-time (heat kernel) technique
\cite{DeWitt65} yields the general expression for the one-loop
divergences in the
form
\beq
\Ga^{(1)}_{div} = - \frac{\mu ^{D-4}}{\vp} \int d^{D}x \sqrt{-g}
\sTr \Big\{ \frac{1}{2}\hat{P}^{2} + \frac{1}{12}\hat{S}_{\mu
\nu}^{2} + \frac{1}{6} \Box \hat{P} + \frac{\hat{1}}{180}
\big(R_{\mu \nu \alpha \beta}^{2} - R_{\mu \nu}^{2} + \Box R \big)
\Big\},
\mbox{\quad}
\label{Sch-DW}
\eeq
where $\vp=(4\pi)^{2}(D-4)$ and
\beq
\hat{P}
&=&
\hat{\Pi} + \frac{\hat{1}}{6}R - \nabla_{\mu} \hat{h}^{\mu}
- \hat{h}_{\mu} \hat{h}^{\mu},
\nn
\\
\hat{S}_{\mu \nu}
&=&
\big[ \nabla_{\nu}, \nabla_{\mu} \big] \hat{1}
+ \nabla_{\nu} \hat{h}_{\mu} - \nabla_{\mu} \hat{h}_{\nu}
+ \hat{h}_{\nu} \hat{h}_{\mu} - \hat{h}_{\mu} \hat{h}_{\nu}.
\label{PS}
\eeq

The relations (\ref{Sch-DW}), (\ref{PS}) lead to the following
result for the one-loop divergences in the model under
consideration\footnote{We present here only the final result,
the intermediate formulas can be found in the Appendix.},
\beq
\Gamma^{(1)}_{div}
&=&
\Ga^{(1)}_{vac,\,div}
\,+\,
\Ga^{(1)}_{m,\,div},
\eeq
where
\beq
\Ga^{(1)}_{vac,\,div}
&=&
- \frac{\mu ^{D-4}}{\vp}
\int d^{D} x \sqrt{-g} \,\bigg\{
\frac12\big( m^4_1 + m^4_2\big) - 2NM^4
\nn
\\
&+&
 \Big( \frac{N}{3}M^2 - m^{2}_1\tilde{\xi}_1
- m^{2}_2\tilde{\xi}_2 \Big)R
+ \Big( \frac{N}{24} +\frac{1}{45}\Big) R_{\mu \nu \alpha \beta}^2
+ \frac{8N-2}{180} R_{\mu \nu}^2
\nn
\\
&+&
\frac12 \Big(\tilde{\xi}_1^2 + \tilde{\xi}_2^2 - \frac{N}{9}\Big)R^2
+ \Big(\frac{1}{45} + \frac{N}{9}
- \frac{1}{6}\tilde{\xi}_1
- \frac{1}{6}\tilde{\xi}_2 \Big)\Box R\bigg\}
\label{Gadivac}
\eeq
and
\beq
\Ga^{(1)}_{m,\,div}
&=&
- \frac{\mu ^{D-4}}{\vp}
\int d^{D} x \sqrt{-g} \,\bigg\{
\sum_k 3\bar{\Psi}_k
\Big[\frac{i}{2} h^{2}_1 \slashed{\nabla}
- \frac{i}{2} h^{2}_2 \slashed{\nabla}
+ h^{2}_1 (M + h_1 \ph - h_2 \ch \ga^5)
\nn
\\
&+&
h^{2}_2 (M+ h_1 \ph - h_2 \ch \ga^5) \Big] \Psi_k
+ 2Nh^{2}_1 (\partial_\mu \ph)^2 -2Nh^{2}_2 (\partial_\mu \ch)^2
\nn
\\
&+&
\Big(\frac13 N h^2_1 - \frac{\la_1}{2}\tilde{\xi}_1
 - \la_3  \tilde{\xi}_2 \Big) R \ph ^2
+ \Big( \frac{1}{8} \lambda^2_1 + \frac12 \la^2_3 -2 N h^4_1 \Big) \ph ^4
\nn
\\
&+&
\Big[\frac{2}{3}Nh_1M - g\tilde{\xi}_1
- p\tilde{\xi}_2
 \Big] R\ph
+ (m^{2}_1 g + m^{2}_2 p - 8Nh_1M^3)\ph
\nn
\\
&+&
\frac{1}{2}\big(g^2 +p^2 +\lambda_1 m^2_1 + 2\la_3 m^2_2
- 24Nh^2_1 M^2\big)\ph ^2
+ \Big( \la_3 p -8N M h^3_1 + \frac12\,g \lambda_1 \Big)\ph ^3
\nn
\\
&+&
\frac{1}{2}\big(2\la_3 m^2_1 + \la_2 m^2_2 + 8p^2
+ 8Nh^2_2 M^2\big)\ch ^2
+ \Big( \frac12 \la^2_3 + \frac18 \la^2_2 -2Nh^4_2 \Big) \ch^4
\nn
\\
&-&
\frac12 \Big( \la_2 \tilde{\xi}_2
+ 2\la_3 \tilde{\xi}_1 + \frac{2N}{3} h^2_2 \Big) R\ch^2
+ \frac12 \big( \la_1 \la_3 + \la_2 \la_3 +8Nh^2_1 h^2_2
+ 32 \la^2_3 \big) \ph^2 \ch^2
\nn
\\
&+&
\Big( g\la_3 + \frac12 p\la_2 + 16p\la_3 + 8Nh_1 h^2_2 M \Big)\ph \ch^2
+ \frac16\,\Big( g + p - 8Nh_1M \Big) \Box \ph
\nn
\\
&+&
\frac{1}{12}\,\Big(\lambda_1 + 2 \la_3 - 16 Nh^{2}_1 \Big)\Box \ph^2
+ \frac{1}{12}\,\Big(\lambda_2 + 2 \la_3 + 16 Nh^{2}_2\Big)\Box \ch^2
\bigg\}.
\label{Gadiv}
\eeq
For compactness, we have introduced the notations
$\tilde{\xi}_{1,2}=\xi_{1,2} - \frac16$. The vacuum
divergences are included for the sake of completeness.

The expression (\ref{Gadiv}) shows that the odd terms, which we
have included in the classical action \eqref{1act}, subject to the
divergences. Exactly as it is the case in the simpler single-scalar
theory, these terms have no symmetry protection and the structure
of divergences is exactly as should be expected from the symmetry
and power-counting arguments.

\section{Renormalization }
\label{s3}

Once the form of the one-loop divergences is known one can easily
find the relations between bare and renormalizable quantities. For the
fields we meet
\beq
\ph_0
&=&
\mu^{\frac{D-4}{2}} \Big(1+ \frac{2Nh^2_1}{\epsilon}\Big) \ph ,
\qquad
\ch_0
\,=\,
\mu^{\frac{D-4}{2}} \Big(1- \frac{2Nh^2_2}{\epsilon}\Big) \ch ,
\nn \\
\Psi_{k0}
&=&
\mu^{\frac{D-4}{2}} \Big[ 1+ \frac{3}{4\epsilon} (h^2_1-h^2_2)\Big]\Psi_k .
\label{fields}
\eeq
The relations for masses have the form
\beq
M_0
&=&
\Big(1-\frac{9}{2\epsilon}h^2_1 - \frac{3}{2\epsilon}h^2_2 \Big) M,
\nn\\
m_{10} ^2
&=&
m^2_1
- \frac{ g^2 + p^2 +4N h^2_1 m^2_1 + \lambda_1 m^2_1
+ 2\la_3 m^2_2 -24N h^2_1 M^2}{\epsilon},
\nn \\
m_{20} ^2
&=&
m^2_2
- \frac{ 8p^2 - 4N h^2_2 m^2_2 + \lambda_2 m^2_2
+ 2\la_3 m^2_1 + 8N h^2_2 M^2}{\epsilon}
.
\label{masses}
\eeq
For the even couplings and nonminimal parameters we find
\beq
\xi_{10}
&=&
\xi_1 -  \frac{\la_1 + 4Nh^2_1}{\epsilon} \, \tilde{\xi}_1
- \frac{2\la_3}{\epsilon} \tilde{\xi}_2 ,
\nn\\
\xi_{20}
&=&
\xi_2 +  \frac{4Nh^2_2 - \la_2}{\epsilon} \, \tilde{\xi}_2
- \frac{2}{\epsilon}\la_3 \tilde{\xi}_1 ,
\nn\\
h_{10}
&=&
\mu ^{\frac{4-D}{2}}
h_1 \Big(1 - \frac{4Nh^2_1 + 9h^2_1 + 3h^2_2}{2\epsilon} \Big) ,
\nn\\
h_{20}
&=&
\mu ^{\frac{4-D}{2}}
h_2 \Big(1 + \frac{4Nh^2_2 + 9h^2_2 + 3h^2_1}{2\epsilon} \Big) ,
\nn\\
\lambda_{10}
&=&
\mu^{4-D}
\Big( \lambda_1
+\frac{48 N h^4_1 -8 N \la_1 h^2_1 - 3\la^2_1 - 12\la^2_3 }{\ep}\Big),
\nn\\
\lambda_{20}
&=&
\mu^{4-D}
\Big(
\lambda_2
+\frac{48 N h^4_2 +8 N \la_2 h^2_2 - 3\la^2_2 -12\la^2_3}{\ep} \Big),
\nn\\
\lambda_{30}
&=&
\mu^{4-D}
\Big( \lambda_3
- \frac{ \la_1 \la_3 + \la_2 \la_3
+ 8 N h^2_1 h^2_2 + 32 \la^2_3 + 4N\la_3 h^2_1
- 4N\la_3 h^2_2}{\ep} \Big).
\label{parameters1}
\eeq
And, finally,  for the odd couplings and nonminimal parameters,
\beq
g_0
&=&
\mu^{\frac{4-D}{2}} \Big(g + \frac{48N M h^3_1 - 3g\lambda_1
-6N h^2_1 g - 6\la_3p}{\epsilon} \Big),
\nn\\
\tau_0
&=&
\mu^{\frac{4-D}{2}}\Big(\tau
+ \frac{8N h_1 M^3 -2N\tau h^2_1 - m^2_1 g - m^2_2 p }{\epsilon}
\Big) ,
\nn\\
p_0 &=&
\mu^{\frac{D-4}{2}} \Big(
p - \frac{1}{\epsilon} \big( 2\la_3g + \la_2p +32\la_3p
+ 16Nh_1 h^2_2 M - 4Nh^2_2 p + 2Nh^2_1p \big) \Big),
\nn\\
f_0
&=&
\mu ^{\frac{D-4}{2}} \Big[ f + \frac{g}{\epsilon}\tilde{\xi}_1
+ \frac{p}{\epsilon} \tilde{\xi}_2
- \frac{2N h_1M + 6N f h^2_1}{3\epsilon} \Big].
\label{parameters2}
\eeq
Note the non-trivial renormalization of the odd coupling parameters
and in particular of the new non-minimal coupling parameter $f$.


The $\beta$- and $\ga$-functions can be calculated from the
renormalization relations for the parameters and fields. For the
theories in curved spacetime the procedure \cite{Buch84,Toms83}
is described in detail in the book \cite{book}, so we give only the
final results for
\beq
\label{betas-def}
\beta_{P} = \lim_{D \to 4} \mu \frac{dP}{d \mu},
\\
 \gamma_{\Phi} \Phi = \lim_{D \to 4} \,\,\mu \frac{d \Phi}{d \mu}
 \n{gamma-def}\,,
\eeq
where $P=\big(m^2_1, m^2_2, M, h_1, h_2, \la_1, \la_2, \xi_1,
\xi_2, g, p, \tau, f\big)$ are the renormalized parameters
and $\Phi= (\ph , \, \ch, \,\Psi_k)$ are the renormalized fields.
Using  the relations \eqref{fields}, (\ref{masses}),
(\ref{parameters1}) and (\ref{parameters2}), we obtain the
following results:
\beq
\beta_{h_1}
&=&
\frac{(4Nh^3_1 + 9h^3_1 + 3h_1 h^2_2)}{2(4\pi)^2} ,
\nn
\\
\beta_{h_2}
&=&
- \frac{(4Nh^3_2 + 9h^3_2 + 3h^2_1 h_2)}{2(4\pi)^2} ,
\nn
\\
\beta_M
&=&
\frac{3 M}{2(4\pi)^2} \Big(3h^2_1 + h^2_2 \Big) ,
\nn
\\
\beta_{\lambda_1}
&=&
\frac{1}{(4\pi)^2} \Big( 3\lambda^2_1 + 8N\lambda_1 h^2_1
- 48N h^4_1 +12 \la^2_3 \Big) ,
\nn
\\
\beta_{\lambda_2}
&=&
\frac{1}{(4\pi)^2} \Big(3 \la^2_2 -8 N\lambda_2 h^2_2
- 48N h^4_2 +12 \la^2_3 \Big) ,
\nn
\\
\beta_{\lambda_3}
&=&
\frac{1}{(4\pi)^2} \Big( \la_1 \la_3 + \la_2 \la_3 + 32 \la_3^2
+ 8 N h_1^2 h_2^2 + 4 N \la_3 h_1^2 -  4 N \la_3 h_2 ^2 \Big),
\nn
\\
\beta_{\xi_1}
&=&
\frac{1}{(4\pi)^2} \Big[
\Big(\la_1 + 4Nh^2_1
\Big) \Big(\xi_1 - \frac{1}{6}\Big)
+ 2\la_3\Big(\xi_2 - \frac{1}{6} \Big) \Big],
\nn
\\
\beta_{\xi_2}
&=&
\frac{1}{(4\pi)^2} \Big[
\Big(\la_2  - 4Nh^2_2
\Big) \Big(\xi_2 - \frac{1}{6}\Big)
+ 2\la_3\Big(\xi_1 - \frac{1}{6} \Big) \Big],
\nn
\\
\beta_g
&=&
\frac{1}{(4\pi)^2} \Big(3 g\lambda_1 + 6 N gh^2_1 - 48N Mh^3_1
+ 6p \la_3\Big) ,
\nn
\\
\beta_p
&=&
\frac{1}{(4\pi)^2} \Big(p\lambda_2 +32 \la_3 p - 4N ph^2_2 + 2N ph^2_1
+ 2g\la_3 + 16NM h_1 h^2_2 \Big) ,
\nn
\\
\beta_{m^2_1}
&=&
\frac{1}{(4\pi)^2}
\Big[m^2_1 \lambda_1 + g^2 + p^2 + \Big(4 m^2_1 - 24M^2 \Big)Nh^2_1
+ 2 \la_3 m^2_2 \Big] ,
\nn
\\
\beta_{m^2_2}
&=&
\frac{1}{(4\pi)^2}
\Big[m^2_2 \lambda_2 + 8p^2 + \Big(8M^2 - 4m^2_2 \Big)Nh^2_2
+ 2 \la_3 m^2_1 \Big] ,
\nn
\\
\beta_{\tau}
&=&
\frac{1}{(4\pi)^2}\Big(2N \tau h^2_1 + g m^2_1 + pm^2_2 -8Nh_1M^3 \Big) ,
\nn
\\
\beta_f
&=&
\frac{1}{(4\pi)^2}
\Big[ 2N f h^2_1 - g \Big(\xi_1 - \frac{1}{6} \Big)
- p \Big(\xi_2 - \frac{1}{6} \Big)+ \frac{2}{3} N Mh_1\Big] \,.
\label{betas}
\eeq
For the $\gamma$-functions we have
\beq
\gamma_\ph \,=\, -\frac{2N h^2_1}{(4\pi)^2},
\qquad
\gamma_\ch \,=\, \frac{2N h^2_2}{(4\pi)^2},
\qquad
\gamma_{\Psi_{k}} \,=\, \frac{3}{4(4\pi)^2}(h^2_2 - h^2_1).
\label{gammas}
\eeq
A good check is that, if considering the conformal invariant theory,
with vanishing masses and other dimensional constants, $g$, $p$, $\tau$
and $f$, and assuming $\xi_1=\xi_2=\frac16$, the pole coefficient in
the expression for the divergences \eqref{Gadiv} is also conformal
invariant. Consequently, the $\be$-functions for  $\xi_1$ and $\xi_2$
in this case are linear combinations of $\tilde{\xi}_1$ and
$\tilde{\xi}_2$, defined after Eq.~(\ref{Gadivac}).

\section{Effective potential}
\label{s4}

In this section we derive the one-loop effective potential in the
model under consideration up to first order in scalar curvature,
using the local momentum representation, based on the Riemann
normal coordinates. This method is quite efficient for mass-dependent
calculations of local quantities, such as the effective potential.

The effective potential $V_{eff} (\ph)$ is defined as the
zeroth-order term in the derivative expansion of the effective
action of a background scalar field $\ph(x)$,
\beq
\Ga \,[\ph,g_{\mu\nu}] =
\int d^Dx \, \sqrt{-g} \, \Big\{ - V_{eff} (\ph)
+ \frac12\, Z(\ph) \, g^{\mu\nu} \pa_{\mu} \ph \, \pa_{\nu} \ph
+ \cdots \Big\} ,
\label{as}
\eeq
where $D$ is the spacetime dimension.

Within the loop expansion of the effective action, the corresponding
one-loop correction to the effective potential is given by
\beq
\n{1loopV}
 \int d^D x \sqrt{-g}\,\,\, V (\ph)
&=&
\frac{1}{2}\, \sTr \ln \hat{H}\,\Big|_{\ph = 
const}
\eeq
where $ \hat{H} $ is the bilinear operator of action \eqref{action}.

The curvature expansion of
${V} (\ph)$ reads
\beq
\label{exH}
{V}  &=& {V}_0  \,+\, {V}_1
\,+\, \cdots\,,
\eeq
where $V^{(1)}_0$ is the well-known flat-spacetime effective
potential, which has been derived many times and in different ways
starting from the work of Coleman and Weinberg~\cite{ColeWein}
and ${V}_1$ is the first order in scalar curvature $\,R$. In curved
spacetime the potential can also be derived in different ways.

Let us emphasize that in all known examples the effective potential
can be obtained by solving the renormalization group equation for
the effective action, in both flat \cite{ColeWein} and curved
\cite{BuchOd84} spacetimes (see e.g. Ref.~\cite{book}
for detailed introduction and further references. The generalization
to the model with a single sterile scalar proceeds is done in a close
analogy
to the standard approach, but with some modification due to the
presence of the odd interaction terms \cite{Barra}.

The renormalization group equation for the the effective action
has the form \cite{Buch84,book}
\beq
\Big\{\,\mu\frac{\pa}{\pa\mu}+\be_P
\,\frac{\pa}{\pa P}&+&\int d^Dx
\,\,\ga_{\Phi}\Phi \,\frac{\de}{\de \Phi (x)}
\,\Big\}
\Ga[g_{\al\be},\Phi,P,D,\mu]=0,
\label{n12}
\eeq
where we assume the sum over all parameters (couplings and
masses) $P$ and the fields $\Phi=(\ph,\,\chi,\,\Psi_k)$. The
effective potential satisfies the same equation, due to the
separation of different terms in (\ref{as}). Then, the result for,
e.g., a single scalar field can be presented as a general
symbolic expression\footnote{We will not write down similar
formula for the theory (\ref{classact}), because it is too long.
The interested reader can easily obtain it by analogy with
Eq.~(\ref{Veff-RG}).}
\beq
V_{eff}&=& - \frac12 \, m^{2} \ph^2
-  \frac12 \, \xi R \ph^2
+ \frac{\la}{4!}\,\ph^4
+  \frac{g}{3!} \ph^3
+  \tau \ph
+  fR \ph
\nn
\\
&-&
\frac{1}{4} \ph^2(\beta_{m^2}+2m^2\gamma_{\ph})
\Big[ \ln \Big( \frac{\ph^2_{1*}}{\mu^2}\Big) + C_1\Big]
-
\frac{1}{4} R\ph^2 (\beta_\xi + 2\xi \gamma_{\ph})
\Big[ \ln \Big( \frac{\ph^2_{2*}}{\mu^2}\Big) + C_2\Big] \nn
\\
&+&
\frac{1}{12} \ph^3 (\beta_g + 3g\gamma_\ph)
\Big[ \ln \Big( \frac{\ph^2_{3*}}{\mu^2}\Big) + C_3\Big]
+
\frac{1}{48}\ph^4 (\beta_\lambda + 4\lambda \gamma_\ph)
\Big[ \ln \Big( \frac{\ph^2_{4*}}{\mu^2}\Big) + C_4\Big]
\nn
\\
&+&
\frac{1}{2} \ph (\beta_\tau + \tau \gamma_\ph)
\Big[ \ln \Big( \frac{\ph^2_{5*}}{\mu^2}\Big) + C_5\Big]
+
\frac{1}{2} R\ph (\beta_f + f\gamma_\ph)
\Big[ \ln \Big( \frac{\ph^2_{6*}}{\mu^2}\Big) + C_6\Big]\,,
\label{Veff-RG}
\eeq
where all beta- and gamma-functions are given in Eqs.~(\ref{betas})
and (\ref{gammas}).
The set of the constants $C_{1\,...\,6}$ in the last expression
(\ref{Veff-RG}) can
be found from the initial renormalization conditions. For instance,
the two well-known values, corresponding to the standard choices
in the massless scalar case are $C_4=-\frac{25}{6}$ obtained in
\cite{ColeWein} and $C_2=-3$ obtained in \cite{BuchOd84,book}.

The symbolic expressions $\ln\big({\frac{\ph_{k*}^2}{\mu^2}}\big)$
with $k=1,2,...,6$, in the formula (\ref{Veff-RG}) depend on the
theory under consideration. For instance, in the model with a single
sterile scalar \cite{Barra}, these quantities appear as linear
combinations of the logarithms
\beq
t^{(0)}\,=\,\frac{1}{2}\ln
\Big[\frac{m^2+\frac{1}{2}\la\ph^2+g\ph}{\mu^2}\Big]
\label{t}
\eeq
and
\beq
t^{(\frac{1}{2})}=\frac{1}{2}\ln
\Big[\frac{(M+h\ph)^2}{\mu^2}\Big].
\label{t-1}
\eeq
for the scalar and fermion  contributions to the effective potential,
correspondingly. Namely, the logarithms (\ref{t}) and (\ref{t-1})
are used as an efficient Ansatz to solve the renormalization group
equation for the effective potential.

In the massless case or in the limit of large-scalar limit, one should
expect that the asymptotic behavior  of all terms should be
\beq
\ln\Big({\frac{\ph_{k*}^2}{\mu^2}}\Big)
\,\, \propto\,\, \ln\Big({\frac{\ph^2}{\mu^2}}\Big).
\label{assim}
\eeq
In the subsequent subsection we perform direct calculation of the scalar contribution in the model (\ref{classact})
and show that the result is inconsistent with the expectation based on the Ansatz that consists in guessing the form of the logarithms, such as (\ref{t}) and (\ref{t-1}).

The calculations presented below were performed in the
covariant cut-off regularization of the Euclidean integrals in
the local momentum representation. In the case of effective
potential this regularization is the simplest options. On the other
hand, the transition to the covariant cut-off in the proper time
integral, and consequently to the dimensional regularization is
automatic, as discussed in \cite{Sobreira} (see also earlier
general investigation in flat spacetime \cite{Liao}).

\subsection{Two-scalar sector}

Let us start from the bilinear form of the action in the scalar
sector of \eqref{action} in the form
\beq
S^{(2)}_0 &=&
\frac{1}{2} \int d^{4}x \sqrt{-g} \,\,\,
\Big( \sigma \quad \rho \Big) \,\hat{H}_s \,
\left(
\begin{array}{c}
\si  \\
\rho
\end{array}
\right) \, ,
\label{S2}
\eeq
where the matrix operator has the form
\beq
\n{Bili}
\hat{H}_s = \Box \hat{1} + \left(
\begin{array}{ccc}
M_{11}^2 & M_{12}^2 \\
M_{21}^2 & M_{22}^2
\end{array}
 \right)
\eeq
with $\,\hat{1} = \diag(1,1)$ and
\beq
M_{11}^2  &=&  \tilde{m}_1^2 - \xi_1 R, \nn \\
M_{22}^2  &=& \tilde{m}_2^2 - \xi_2 R , \nn \\
M_{12}^2  &=& M_{21}^2 = 2 p \ch + 4\la_3 \ph \ch
\,,
\eeq
where
\beq
\tilde{m}_1^2 &=& m_1^2 + \frac{\la_1}{2} \ph^2 + \la_3 \ch^2 + g\ph \, ,
\\
\tilde{m}_2^2  &=& m_2^2 + \frac{\la_2}{2} \ch^2 + \la_3 \ph^2 + p\ph \, .
\eeq

In order to simplify the calculations let us diagonalize the matrix
in the second term of relation \eqref{Bili}, by making a rotation in
the space of the fields,
\beq
\left( \begin{array}{c}
\si  \\
\rho
\end{array}
\right) =
U
\left(
\begin{array}{c}
\phi_1  \\
\phi_2
\end{array}
\right)
\,,\quad
\mbox{where}
\quad
U = \left(
\begin{array}{cc}
\cos{ \al} & - \sin{\al} \\
\sin{\al} & \cos{\al}
\end{array}
\right)\,.
\label{rot}
\eeq
After this transformation,  Eq.~\eqref{S2} becomes
\beq
S^{(2)}_0
& = &
\frac{1}{2} \int d^{4} x \sqrt{-g} \,  \Bigg\{  \phi_1 \Box \phi_1
+ \phi_2 \Box\phi_2
\nn
\\
&+&
\phi_1  \left[
\cos^2(\al) M_{11}^2 + \sin^2(\al) M_{22}^2 -  \sin(2\al)  M_{12}^2
\right] \phi_1
\nn
\\
&+&
\phi_2\left[ \sin^2(\al)
 M_{11}^2 + \cos^2(\al) M_{22}^2
 + \sin(2\al)  M_{12}^2
\right] \phi_2
\nn \\
&+&
\phi_1 \left[ \sin(2\al)  \left(M_{22}^2 -  M_{11}^2 \right)
 + 2 \cos(2\al)  M_{12}^2
\right] \phi_2
 \Bigg\}.
\n{S2t}
\eeq
Now, we can simply choose
\beq
\cos(2\al) = \frac{M_{22}^2 -  M_{11}^2 }{2 M_{12}^2} \sin(2\al)
\quad
\Longrightarrow
\quad
\cot(2\al) = \Th \,=\,\frac{M_{22}^2 -  M_{11}^2 }{2 M_{12}^2} \, ,
\eeq
such that the last term in \eqref{S2t} vanishes and the new
diagonal matrix $ \hat{\mathcal{H}}_s = U^{-1}\hat{H}_s U $
has the form
\beq
\n{Hdiag}
\hat{\mathcal{H}}_s  = \Box \hat{1} + \left(
\begin{array}{ccc}
a M_{11}^2 + b M_{22}^2 - c  M_{12}^2 & 0\\
0 & b
 M_{11}^2 + a M_{22}^2
 + c  M_{12}^2
\end{array}
 \right) ,
\eeq
where
\beq
a  =  \frac12 + \frac{\Th}{2 \sqrt{1+\Th^2}}, \qquad
b  = \frac12 -  \frac{\Th}{2 \sqrt{1+\Th^2}}, \qquad
c =  \frac{1}{\sqrt{1+\Th^2}} \, .
\eeq

Since we are interested in the ${\cal O}(R)$-approximation,
it is useful to rewrite \eqref{Hdiag} as
\beq
\n{Hs}
\hat{\mathcal{H}}_s =
\left(
\begin{array}{ccc}
 \Box  - \Pi_1 - \ze_1 R +  (R^2 \cdots) & 0  \\
0 &  \Box  - \Pi_2 - \ze_2  R + O (R^2 \cdots)
\end{array}
\right),
\eeq
with
\begin{align}
&
\Pi_1 = - \left( \frac12+  \frac{\Th_0}{2 \sqrt{1+\Th_0^2}} \right) \tilde{m}_1^2 - \left( \frac12-  \frac{\Th_0}{2 \sqrt{1+\Th_0^2}} \right)\tilde{m}_2^2
 + \frac{M_{12}^2}{\sqrt{1+\Th_0^2}}
\, ,
\\ &
\Pi_2 = - \left( \frac12-  \frac{\Th_0}{2 \sqrt{1+\Th_0^2}} \right) \tilde{m}_1^2 - \left( \frac12+ \frac{\Th_0}{2 \sqrt{1+\Th_0^2}} \right)\tilde{m}_2^2
 - \frac{M_{12}^2}{\sqrt{1+\Th_0^2}} ,
\end{align}
and
\beq
\ze_1 &=& \left( \frac12+  \frac{\Th_0}{2 \sqrt{1+\Th_0^2}} \right) \xi_1
+ \left( \frac12 -  \frac{\Th_0}{2 \sqrt{1+\Th_0^2}} \right) \xi_2 ,
 \\
 \ze_2  &=&  \left( \frac12-  \frac{\Th_0}{2 \sqrt{1+\Th_0^2}} \right) \xi_1
+ \left( \frac12 +  \frac{\Th_0}{2 \sqrt{1+\Th_0^2}} \right) \xi_2 ,
 \eeq
where we denote $\Th_0 = \frac{ \tilde{m}_2^2 - \tilde{m}_1^2}{2 M_{12}^2} $.

As next step, we define
\beq
\n{H1H2}
\hat{\mathcal{H}}_{s}  =
\left(
\begin{array}{cc}
H^{(1)} & 0 \\
0 & H^{(2)}
\end{array}
\right) ,
\eeq
where $H^{(1)} = \left( \Box  - \Pi_1 - \ze_1 R \right)$ and
$ H^{(2)} =\left( \Box  - \Pi_2 - \ze_2 R\right)$,  so that
\beq
\n{DetH}
\Tr \ln \hat{\mathcal{H}}_{s} =
 \ln  \Det \hat{\mathcal{H}}_{s} =
\Tr \ln   H^{(1)} + \Tr \ln  H^{(2)}  \, .
\eeq
Let us point out that using the rotation in the fields space to
diagonalize $ \hat{H} $, we also have to consider the contribution
that comes from the transformation
$\,\hat{\mathcal{H}}_{s} =  U^{-1}  \hat{H}_s  U$.
In relation \eqref{DetH}, since $ \Det U = \Det U^{-1} = 1$, then
there is no contribution. However the Jacobian of a such
transformation in four-dimensional spacetime is proportional to
$\,\de^{4}(0) $, which in dimensional regularization formally
vanishes while in our case via cut-off regularization scheme
this means a cut-off dependent contribution to the cosmological
constant.

Another important observation is that rotation (\ref{rot}) and
an expansion to the first order in curvature are not commuting
operations. This means that if we extract the ${\cal O}(R)$-term
first and after that make a rotation only for a flat-space sector,
the result would be different and not satisfactory from the point
of view of our calculations.

Starting from this point, we meet a product of two normal scalar
operators \eqref{H1H2} and it is possible to use the technique
elaborated in \cite{BP} (see also \cite{Sobreira, Reis, Barra}) to
find the one-loop effective potential up to first order in $\,R$,
using the Riemann normal coordinates formalism.

The equation for the propagator of a scalar field $G_c (x,x')$
related to $ H^{(1)} $ has the form
\beq
\n{GH1}
\left(g^{\frac14} \Box g^{-\frac14} - \Pi_1 - \ze_1 R
\right)
\bar{G} (x,x') \,=\,-\, \de^D (x - x')\,.
\eeq
In Eq.~\eqref{GH1} we take into account the expression for
the covariant Dirac delta function
\beq
\de_c (x,x') = g^{-1/4} \, \de^D (x - x') \,g'^{\,-1/4}
\eeq
and the modified propagator $\bar{G} (x,x')$~\cite{BP}. Both
elements are necessary for the consistency of the expansion, so that the {\it r.h.s.} of the above equation does not depend on the metric tensor.  Thus, one can use the relation
$\,\Tr \ln \hat{H} = - \,\Tr \ln \bar{G} $ to derive the
dependence on the curvature tensor in Eq.~\eq{1loopV}.

In the Riemann normal coordinates the expansion of the
spacetime metric $ g_{\al\be}$ up to first order in the
curvature is given by \cite{Petrov} (see also simplified
introduction and more references in \cite{Tensors})
\beq
g_{\al\be}(x) &=&\eta_{\al\be}
- \frac{1}{3}\,R_{\al\mu\be\nu}(x') \,y^\mu y^\nu
+ \cdots \,,
\eeq
hence
\beq
R(x) &=& R(x') + \cdots , \label{eR}
\\
\square &=& \pa^2
+ \frac{1}{3}\, R^{\mu\,\nu}_{\,\al\,\be}(x')
\,y^\al y^\be \, \pa_\mu \pa_\nu
-\frac{2}{3}\,R^{\al}_{\be}(x') \, y^\be \, \pa_\al
+ \cdots\,\,.
\label{ebox}
\eeq
Starting from this formula, it is easy to get
\beq
g^{1/4} \, \square \, g^{-1/4} &=&
 \pa^2 + \frac16\, R
+ \frac{1}{3}\left( R^{\mu\,\nu}_{\,\al\,\be} (x')
\,y^\al y^\be \, \pa_\mu \pa_\nu
- R^{\al}_{\be} (x') \, y^\be \, \pa_\al \right) + \cdots\,,
\label{eboxg}
\eeq
where the derivatives are $\pa_\al = \frac{\pa}{\pa y^\al}$,
$\,\pa^2 = \eta^{\mu\nu} \pa_\mu \pa_\nu\,$ and the dots
mean higher order terms in the curvature tensor and its
covariant derivatives.

After all, Eq.~\eqref{GH1} becomes
\beq
\n{G}
\Big[
-\pa^2 + \Pi_1 + \Big( \ze_1 -\frac{1}{6} \Big) R \Big]
\bar{G} (x,x') \,=\, \de^D (x - x')\,.
\eeq
We can also note that the last term in Eq.~\eqref{eboxg} does not contribute to the effective potential due to the Lorentz invariance \cite{BP}.

The solution up to the first order in the curvature has the form
\beq
\n{BPs}
\bar{G}(x,x')
&=& \int \frac{d^D k}{(2\pi)^D}
\, e^{iky} \left[\frac{1}{k^2 + \Pi_1}
- \Big( \ze_1 -\frac{1}{6} \Big)
\frac{R}{(k^2 + \Pi_1)^2}
\right] .
\eeq

The results presented above, enable one to find the one-loop
effective potential. Taking into account the expansion of bilinear
operator $ {H}^{(1)} $ and the Green's function $\bar{G}(x,x')$
up to first order in $\,R$, we have
\beq
\label{trG}
 - \Tr \ln \bar{G}
\,=\,
 \Tr \ln (\hat{H}^{(1)}_0 + \hat{H}^{(1)}_1 R)
\,=\,
\Tr \ln \hat{H}^{(1)}_0 + \Tr \bar{G}_0 \hat{H}^{(1)}_1 R .
\eeq

For the effective potential in flat spacetime we need just the
first term in the \emph{r.h.s.} of  Eq.~\eq{trG} given by
\beq
\label{V0D}
- \int d^D x \, \bar{V}^{(1)}_0  &=&
\frac{1}{2}\Tr \ln \hat{H}^{(1)}_{0}
=  \frac{1}{2} \Tr \ln S_2(\ph, \ch)
- \frac{1}{2} \Tr \ln S_2(\ph = \ch = 0),
\eeq
where $S_2(\phi)$ is the bilinear form of the classical action
in the background-field formalism. The last term in Eq.~\eq{V0D}
can be seen as a normalization of the functional integral. This
term arises naturally through the diagrammatic representation
of effective potential. From Eq.~\eq{V0D} we get
\beq
\bar{V}_0^{(1)}(\ph,\ch) \,=\,
\frac12 \int \frac{d^D k}{(2\pi)^D} \,
\ln \Big( \frac{k^2+ \Pi_1}{k^2+m_1^2 }\Big)\,.
\eeq
Using the Euclidean momentum cut-off $ \Om $, for $D = 4$
we have
\beq
\bar{V}_0^{(1)}(\ph, \ch) \,=\,
\frac{1}{ 2(4\pi)^2}
\int_0^{\Om} dk^2 \, k^{2} \ln
\Big( \frac{k^2+ \Pi_1}{k^2+m_1^2 }  \Big),
\eeq
and we finally get
\beq
\bar{V}^{(1)}_0(\ph,\ch)  =
\bar{V}^{(1)}_{0-div}(\ph, \ch) + \bar{V}^{(1)}_{0-fin}(\ph, \ch) ,
\eeq
where
\beq
\label{V0}
\bar{V}^{(1)}_{0-div}(\ph, \ch)
&=&
\frac{1}{32\pi^2} \bigg\{
\Om^2 \left( \Pi_1 - m_1^2  \right)
- \frac{\Pi_1^2 }2 \ln \frac{\Om^2}{\mu^2}
+
\frac12 m_1^4 \ln \frac{\Om^2}{\mu^2}
\bigg\},
\\
\bar{V}^{(1)}_{0-fin}(\ph , \ch)
&=&
\frac{1}{32\pi^2}
\bigg\{
- \frac14 \left( \Pi_1^2 - m_1^4 \right) + \frac{\Pi_1^2 }2 \ln \frac{\Pi_1}{\mu^2}
-
\frac12 m_1^4 \ln{\frac{m_1^2}{\mu^2}} \bigg\}\,.
\eeq

The second term in the \emph{r.h.s.} of Eq.~\eq{trG} corresponds
to the first order in curvature correction $\bar{V}^{(1)}_1(\ph, \ch)$,
which can be derived as follows
\beq
\label{GHR}
 -\int d^D x \, \bar{V}^{(1)}_1  &=&
 \frac{1}{2} \Tr \bar{G}_{0} \, \hat{H}_1 R
\,=\, - \, \frac{1}{2}  \int d^D x \int d^D x'\,
\bar{G}^{-1}_{0}(x,x') \, \bar{G}_{1}(x',x) R,
\mbox{\quad}
\eeq
so that
\beq
\n{GHR1}
\bar{V}^{(1)}_1 &=&
\frac{1}{2} \int d^D x'\,
\int \frac{d^D k}{(2\pi)^D} \, e^{ik(x-x')}
\int \frac{d^D p}{(2\pi)^D} \, e^{ip(x'-x)}  \,
\bar{G}^{-1}_{0}(k) \, \bar{G}_{1}(p) \, R
\nonumber \\
&=&
\frac{1}{2} \int \, \frac{d^D k}{(2\pi)^D} \,
\bar{G}^{-1}_{0}(k) \, \bar{G}_{1}(k) \, R \,.
\eeq
For $D = 4$ and replacing the explicit forms of
$\bar{G}^{-1}_{0}(k)$  and  $\bar{G}_{1}(k)$ of
Eq.~(\ref{BPs}) in Eq.~\eq{GHR1}, one arrives at
\begin{eqnarray}
\n{v1}
\bar{V}^{(1)}_1
\,=\, -\,
 \frac{1}{2 (2\pi)^4}\,\Big( \ze_1-\frac{1}{6} \Big)R
\int_0^{\Om} \frac{k^2 d k^2}{k^2 + \Pi_1}\, .
\end{eqnarray}
After taking the last integral, the result has the form
\beq
\bar{V}^{(1)}_1(\ph, \ch)  =
\bar{V}^{(1)}_{1-div}(\ph , \ch) + \bar{V}^{(1)}_{1-fin}(\ph, \ch) ,
\eeq
where
\begin{align}
&
\label{V1}
\bar{V}^{(1)}_{1-div}(\ph, \ch) \,=\, -\,
 \frac{1}{32 \pi^2} \Big( \ze_1-\frac{1}{6} \Big) \, R
\left[
\Omega^2 -  \Pi_1 \ln{\frac{\Omega^2}{ \mu^2}}
\right],
\nn \\ &
\bar{V}^{(1)}_{1-fin}(\ph, \ch)  \,=\, -\,
\frac{1}{32 \pi^2} \Big( \ze_1-\frac{1}{6} \Big)\, R \,
 \Pi_1 \ln{ \frac{\Pi_1}{\mu^2}}\,.
\end{align}

We have described the calculations for the first contribution
due to $ \hat{H}^{(1)} $. For the second term $ \hat{H}^{(2)}$
the calculations are analogous except that in this case we have to
use $\Pi_2$ and $\ze_2 $ instead $\Pi_1$ and $\ze_1$.
The final result has the form
\begin{align}
\bar{V}(\ph, \ch)  = & \,
\bar{V}^{(1)}_0(\ph, \ch) + \bar{V}^{(2)}_0(\ph, \ch) +
\bar{V}^{(1)}_{1}(\ph , \ch) + \bar{V}^{(2)}_{1}(\ph, \ch)
\nn \\  = & \,
\frac{1}{32\pi^2}
\bigg\{
- \frac14 \left[ \Pi_1^2 + \Pi_2^2 - (m_1^4+m_2^4) \right]
\nn \\ &
+ \Om^2 \left[ \Pi_1 + \Pi_2 - (m_1^2+ m_2^2) \right]
\nn \\ &
+ \frac12 (\Pi_1^2+\Pi_2^2 )\ln{ \frac{\Pi_2}{\mu^2}}
- \frac12 (\Pi_1^2 +\Pi_2^2)\ln{ \frac{\Om^2}{\mu^2}}
\nn \\ &
- \frac12 m_1^4 \ln{\frac{m_1^2}{\mu^2}}
- \frac12 m_2^4 \ln{\frac{m_2^2}{\mu^2}}
+ \frac12 (m_1^4+m_2^4) \ln \frac{\Om^2}{\mu^2}
\nn \\ &
-\, \Big( \ze_1-\frac{1}{6} \Big) R
\bigg[\Pi_1 \ln{\frac{\Pi_1}{\mu^2}} + \Omega^2 - \Pi_1 \ln{\frac{\Omega^2}{\mu^2}}
\bigg]
\nn \\
&
-\,
\Big( \ze_2-\frac{1}{6} \Big) R
\bigg[
\Pi_2 \ln{ \frac{\Pi_2}{\mu^2}}
+ \Omega^2 - \Pi_2 \ln{\frac{\Omega^2}{\mu^2}}
\bigg]
\, \bigg\} \, .
\label{scalar_contribution}
\end{align}
This is the final result for scalar fields loop to effective potential.

Some observations concerning the expression
(\ref{scalar_contribution}) are in order. First of all, the divergent
part is in the perfect correspondence with the corresponding
part of the result  (\ref{Gadiv}), obtained on the base of the
heat-kernel method. In order to see this it is sufficient to use
the well-known correspondence between covariant cut-off
and dimensional regularization parameter (see, e.g., \cite{bavi85}),
\beq
\frac{2}{4-n}\,\mu^{n-4}
\,\,\sim\,\,\ln \frac{\Om^2}{\mu^2}, \qquad n \longrightarrow 4.
\eeq

Second, the dependence on the renormalization parameter
$\mu$ is exactly the standard one, such that the effective
potential is a solution of the standard renormalization group
equation (\ref{n12}).

Thus, the expression (\ref{scalar_contribution}) indicates that the
quantum corrections are given by some logarithmic terms, similar to
the general renormalization group - based form (\ref{Veff-RG}). On
the other hand, the logarithmic terms in (\ref{scalar_contribution})
depend on the unusual arguments representing the mixture of
different scalar fields, their masses and coupling constants. This
situation is in fact typical for the quantum corrections coming from
the loops with mixed internal lines, e.g. of the light and heavy
mass fields \cite{Ilisie} (see also a recent work \cite{324} for the
extension to curved space). However, it is interesting to
point out that this form of the effective potential does not
confirm a naive expectation that the scalar fields contribution to
effective potential can be obtained using the anzatz of the
form $t^{(0)}$ from Eq.~(\ref{t}) for each of the background
scalars. This output means that the possibility to derive the full
result (\ref{scalar_contribution}) from the renormalization group
equation is not evident and deserves a further study.

We can point out that in the limit of large scalar fields, when
both $|\ph| \to \infty\,$ and $\,|\chi| \to \infty$, our result
(\ref{scalar_contribution}) reduce to the sum of logarithmic
contributions of the scalar fields. However, in general the
effective potential has more complicated form.
The origin of this feature of
the two-scalar model is the rotation (\ref{rot}), that mixes
different masses, interactions and non-minimal parameters.

\section{Conclusions and Perspectives}
\label{s6}

We have formulated the Yukawa model of one sterile scalar and one
axial scalar (pseudoscalar) fields, interacting to themselves and
also to the set of fermions through the Yukawa couplings.

The power counting analysis of the divergences shows that the
helps us to identify the form of the classical potential of scalar and
pseudoscalar self-interaction, providing renormalizable quantum
theory. This potential has all even and odd terms that are allowed
by symmetries (including parity) of the classical theory, without
coupling constants with the inverse-mass dimensions.

The complete analysis of one-loop renormalization, $\beta-$ and
$\gamma$-functions was given in Sec.~\r{s3}. The main results
of this part is the importance of the mixed scalar-pseudoscalar
terms, which do not have symmetry protection and, as a result, are
indispensable for renormalizability of the theory. Thus, we have
completely described the one-loop renormalization structure of
the model under consideration.

The effective potential has been calculated up to the linear in
scalar curvature terms. The results is a sum of independent
contributions from the scalar fields loop and from the spinor
field loop. The contributions of the scalar sector has been 
calculated in the explicit form and demonstrate a nontrivial 
dependence on the background scalar fields, on masses and 
coupling constants. Let us note that the derivation
of the fermion contribution to effective potential in the full
massive theory faces serious technical difficulties and we left
it for the future work.

It is interesting that unlike the single scalar field models, the
effective potential in the two-scalar model under discussion
contains usual logarithmic terms and also the terms with the
non-logarithmic asymptotic. In the scalar loop sector the
model under consideration is qualitatively similar to the
situation with two quantum fields with different masses,
that is well-known from the literature (see, e.g. \cite{Ilisie})
and was recently discussed in curved space \cite{324}.
It is remarkable, however, that in our expressions we could
observe the effect of masses even in the local effective
potential, without invoking the non-local form factors, as
it is done in the mentioned publications. It is worth mentioning,
that the direct calculation of the scalar loop for effective
potential has been performed using rotation in the space of
the scalar fields. This operation turns out to be not
commutative with the expansion using local momentum
representation.

\section*{Acknowledgments}

E.A.R. is grateful to Coordena\c{c}\~ao de Aperfei\c{c}oamento de
Pessoal de N\'{\i}vel Superior - CAPES  for supporting his Ph.D. project.
I.B. is grateful to CAPES for supporting his
long-term visit to UFJF and to the Physics Departament of UFJF for
kind hospitality during the period when this work started.
Also he thanks the Russian Ministry of Science and
High Education, project No 3.1386.2017 for partial support.
The work of I.L.Sh. was partially supported by Conselho Nacional de
Desenvolvimento Cient\'{i}fico e Tecnol\'{o}gico - CNPq under the
grant 303635/2018-5 and Funda\c{c}\~{a}o de Amparo \`a Pesquisa
de Minas Gerais - FAPEMIG under the project APQ-01205-16.

\section*{Appendix}

The intermediate expressions leading to \eqref{Gadiv} are
\beq
&&
\nabla_{\mu} \hat{h}^{\mu} = \left(
\begin{array}{ccc}
0 & 0 & \frac{i}{2} h_1 \nabla_{\mu} \bar{\Psi}_j \gamma^{\mu}
\\
0 & 0 & \frac{i}{2} h_2 \nabla_{\mu} \bar{\Psi}_j \gamma^5 \gamma^{\mu}
\\
0 & 0 & \frac{i}{2} (h_1 \nabla_{\mu} \ph
+ h_2 \nabla_{\mu} \ch \gamma^5)\gamma^{\mu} \delta^{ij}
\end{array}
\right) ,
\nn
\\
&&
\qquad
\hat{h}_{\mu} \hat{h}^{\mu} =
\left(
\begin{array}{ccc}
0 & 0 &  - h^{2}_1 \bar{\Psi}_k \ph + h_1 h_2 \bar{\Psi}_k \ch \gamma^5
\\
0 & 0 &  - h_1 h_2 \bar{\Psi}_k \gamma^5 \ph + h^{2}_2 \bar{\Psi}_k \ch
\\
0 & 0 & (- h^{2}_1 \ph^{2} + h^{2}_2 \ch^2) \delta^{ik}
\end{array}
\right)
\nn
\\
&& \mathrm{and}
\qquad
\hat{h}_{\mu} \hat{h}_{\nu} = \left(
\begin{array}{ccc}
0 & 0 & -\frac{1}{4} (h^{2}_1 \bar{\Psi}_k \ph
- h_1 h_2 \bar{\Psi}_k \ch \ga^5 ) \gamma_{\mu} \gamma_{\nu} \delta^{ik}
\\
0 & 0 & -\frac{1}{4} (h_1 h_2 \bar{\Psi}_k \ph \ga^5
- h^2_2 \bar{\Psi}_k \ch) \gamma_{\mu} \gamma_{\nu} \delta^{ik}
\\
0 & 0 & - \frac14 (h^2_1 \ph^2 - h^2_2 \ch^2) \ga_{\mu} \ga_{\nu}
\end{array}
\right).
\eeq
As a result, the elements of the matrices ${\hat P}$
and ${\hat S}_{\mu\nu}$ in Eq.~(\ref{PS}) have the form
\beq
P_{11} &=& m^{2}_1 + \frac{\la_1 \ph^2}{2} + g \ph
- \left(\xi_1 - \frac{1}{6} \right) R + \la_3 \ch^2 ,
\nn
\\
P_{12} &=& 2p \ch + 4\la_3 \ph \ch ,
\qquad
P_{21}\,=\, 2p \ch + 4\la_3 \ph \ch ,
\nn
\\
P_{13} &=& h_1 \bar{\Psi}_k (M + h_1 \ph - h_2 \ch \ga^5)
- \frac{i}{2} h_1 (\nabla_{\mu} \bar{\Psi}_k) \gamma^{\mu} ,
\nn
\\
P_{22} &=& m^{2}_2 + \frac{\la_2 \ch^2}{2} + p \ph
- \left(\xi_2 - \frac{1}{6} \right) R + \la_3 \ph^2 ,
\nn
\\
P_{23} &=& h_2 \bar{\Psi}_k (M\ga^5 + h_1 \ph \ga^5 - h_2 \ch)
- \frac{i}{2} h_2 (\nabla_{\mu} \bar{\Psi}_k) \ga^5 \gamma^{\mu} ,
\nn
\\
P_{31} &=& 2h_1 \Psi_i \, ,
\qquad
P_{32} \,=\, 2h_2 \ga^5 \Psi_i \, ,
\nn
\\
P_{33} &=& \Big [ M^{2} -\frac{1}{12} R + h_1 M \ph
+ h_2 M \ch \ga^5 + h^{2}_1 \ph^2 - h^{2}_2 \ch^2
\nn
\\
&-& \frac{i}{2} (h_1 \nabla_{\mu} \ph + h_2 \nabla_{\mu}\ch \ga^5)
\ga^{\mu} \Big] \delta^{ik}
\eeq
and
\beq
S_{\mu \nu \;13} &=&
-\frac{i}{2} h_1 \left[ (\nabla_{\mu}\bar{\Psi}_k) \gamma_{\nu}
- (\nabla_{\nu} \bar{\Psi}_k) \gamma_{\mu} \right]
+ \frac{1}{4} (h^{2}_1 \bar{\Psi}_k \ph
- h_1 h_2 \bar{\Psi}_k \ch \ga^5)
\left[ \gamma_{\mu}, \gamma_{\nu} \right],
\nn
\\
S_{\mu \nu \;23} &=&
-\frac{i}{2} h_2 \left[ (\nabla_{\mu} \bar{\Psi}_k)\ga^5 \gamma_{\nu}
- (\nabla_{\nu} \bar{\Psi}_k)\ga^5 \gamma_{\mu} \right]
+ \frac{1}{4} (h_1 h_2 \ph \bar{\Psi}_k \ga^5
- h^{2}_2 \bar{\Psi}_k \ch) \left[ \gamma_{\mu}, \gamma_{\nu} \right] ,
\nn
\\
S_{\mu \nu \;33} &=&
\Big[ [\nabla_{\nu} , \nabla_{\mu}]
- \frac{i}{2} h_1 \left[ (\nabla_{\mu} \ph) \gamma_{\nu}
- (\nabla_{\nu} \ph) \gamma_{\mu} \right]
-\frac{i}{2} h_2 \left[ (\nabla_{\mu} \ch) \ga^5 \gamma_{\nu}
- (\nabla_{\nu} \ch) \ga^5 \gamma_{\mu} \right]
\nn
\\
&+& \frac{1}{4} (h^{2}_1 \ph^{2} - h^{2}_2 \ch^2)
\left[ \gamma_{\mu}, \gamma_{\nu} \right] \Big] \delta ^{ik} .
\eeq


\end{document}